\documentclass[12pt,a4paper]{article}
\usepackage{graphics}
\newcommand\herw{\small HERWIG}
\newcommand\isajet{\small ISAJET}
\def \lsim            
{\raisebox{-3pt}{$\>\stackrel{<}{\scriptstyle\sim}\>$}}

\begin{document}
\begin{titlepage}
\begin{flushright}
  CERN-TH/2000-116\\
  RAL-TR-2000-006 \\
  UR-1602 \\
  hep-ph/0004179
\end{flushright}
\par \vspace{10mm}
\begin{center}
{\Large \bf \sc
Jet Activity in $t\bar t$ Events \\[1ex]
and Top Mass Reconstruction\\[2ex]
at Hadron Colliders}
\end{center}
\par \vspace{2mm}
\begin{center}
{\bf G. Corcella $^{1}$, M.L. Mangano $^{2}$ and M.H. Seymour $^3$}\\
\vspace{2mm}
{$^1$ Department of Physics and Astronomy, University of Rochester,\\
Rochester, NY 14627, U.S.A.}\\
\vspace{2mm}
{$^2$ CERN, Theory Division,}\\
{CH--1211, Geneva 23, Switzerland\@.}\\
\vspace{2mm}
{$^3$ Rutherford Appleton Laboratory, Chilton,}\\
{Didcot, Oxfordshire.  OX11 0QX\@.  U.K.}
\end{center}

\par \vspace{2mm}
\begin{center} {\large \bf Abstract} \end{center}
\begin{quote}
  \pretolerance 10000 We analyse the impact of matrix-element
  corrections to top decays in \herw\ on several observables related
  to the jet activity in $t\bar t$ events at the Tevatron and at the
  LHC. In particular, we study the effects induced by the higher-order
  corrections to top decays on the top mass reconstruction using the
  recently proposed $J/\psi+\ell$ final states.
\end{quote}

\vspace*{\fill}
\begin{flushleft}
  CERN-TH/2000-116\\
  UR-1602       \\
  RAL-TR-2000-006 \\
  April 2000
\end{flushleft}
\end{titlepage}

\section{Introduction}
Top-quark and in general heavy-flavour production physics (see, for a  
review,
[\ref{frixione}]) is currently one of the main fields of investigation
in both experimental and theoretical particle physics.  At the next
Run~II of the Tevatron accelerator and, in the future, at the
LHC~[\ref{Beneke:2000hk}] and at $e^+e^-$ Linear
Colliders~[\ref{Accomando:1998wt}],  
the production of a large amount of
$t\bar t$ pairs will allow accurate studies of the top quark
properties and an improved measurement of its mass.

The measurement of these properties will rely by-and-large on the
accuracy of the theoretical modeling of the exclusive properties of
the final states, for example jet distributions. Use of inclusive
parton-level calculations, although resulting in the exact inclusion
of the next-to-leading-order (NLO)
contributions~[\ref{Mangano:1992jk}], is not sufficient to fully
describe the effects associated with the large logarithms
corresponding to soft or collinear parton emission. Calculations based
on the fragmentation function formalism~[\ref{Cacciari:1998it}], where
collinear logarithms can be resummed up to the NLO, are on the other
hand too inclusive to allow a complete study of the final states.
Monte Carlo event generators
[\ref{herwig},\ref{pythia},\ref{Paige:1981fb}] are the best possible
tool to perform the resummation of these enhanced logarithms, to
simulate multiple radiation in high energy processes, and to provide a
description of the hadronization transition leading to the final
observable particles.

Inclusion of these higher-order corrections by the QCD event
generators is done however in the soft/collinear approximation.
Furthermore, the Monte Carlo evolution suppresses entirely emission of
radiation inside some regions of the physical phase space (`dead
zones') corresponding to hard and large-angle parton radiation. These
regions are unfortunately sometimes crucial from the experimental
point of view. Emission inside these dead zones can be performed using
the exact amplitudes by following the method discussed in
[\ref{sey1}].  This method has been applied in the past to jet
production in $e^+e^-$ annihilation [\ref{sey2}], in deep inelastic
scattering [\ref{sey3}] and, more recently, to the description of top
decays [\ref{corsey1}] and of Drell--Yan processes [\ref{corsey2}].
Alternative approaches have also been proposed in the literature, see
e.g. ref.~[\ref{mepapers}]. 

In [\ref{corsey1}] a marked impact of matrix-element corrections to
top decays was found for $e^+e^-$ interactions slightly
above the threshold for $t\bar t$ production.  In this paper we wish
to perform a similar analysis for top production and decay at hadron
colliders and investigate the effect of the implemented hard and
large-angle gluon radiation on jet observables and on the top mass
reconstruction.  As far as the top mass is
concerned, we shall consider final states with leptons and $J/\psi$
since the LHC experimentalists claim it is a favourite channel, with a
systematic error no larger than 1 GeV [\ref{avto}], and we shall give
more details on the analysis and the preliminary results presented in
[\ref{corcella}].

In Section 2 we briefly review the method applied in [\ref{corsey1}]
to implement matrix-element corrections to the \herw\ description of
top decays. In Section 3 we shall study
phenomenologically-relevant jet observables at the Tevatron and at the
LHC and investigate the impact of the improved treatment of top
decays. In Section 4 we shall discuss the method of reconstructing the
top mass by using final states with leptons and $J/\psi$ and
the effect of matrix-element corrections on the top mass measurement.
In Section 6 we shall make some concluding remarks and comments on
possible further improvements of the study here presented.

\section{Matrix-element corrections to simulations of top decays}
In the \herw\ Monte Carlo event generator, the top quark decay $t\to
bW$ is performed in the top rest frame, as discussed in
[\ref{marweb1}].  The top quark cannot emit soft gluons in its decay
stage as it is at rest, while the $b$ quark is allowed to radiate in
the cone $0<\theta_g<\pi/2$, $\theta_g$ being the soft-gluon emission  
angle
relative to the direction of the $b$ quark. 
The subsequent parton shower is performed following the
prescription of the angular ordering [\ref{marweb2}].  The $W$
hemisphere, corresponding to $\pi/2<\theta_g<\pi$, is completely empty
and the soft phase space is not therefore entirely filled by the
\herw\ algorithm.  In [\ref{marweb1}] the authors showed that
neglecting the `backward' gluon radiation correctly predicts the total
energy loss, however problems are to be expected when dealing with  
angular
distributions.

In [\ref{corsey1}] matrix-element corrections to the \herw\ simulation
of top decays have been implemented: the missing phase space is
populated according to a distribution obtained from the calculation of
the exact first-order matrix element (hard correction), the shower in
the already-populated region is corrected by using the exact
amplitude any time a hardest-so-far emission is encountered in the
evolution (soft correction).

Since the \herw\ dead zone includes part of the soft singularity,
matrix-element corrections to top decays are not a straightforward
extension of the method applied in
[\ref{sey2},\ref{sey3},\ref{corsey2}].  The soft singularity has been
avoided by setting a cutoff on the energy of the gluons which are
radiated in the backward hemisphere by the $b$ quark.  As shown in
[\ref{corsey1}], and discussed later on in this paper, the
sensitivity to this cutoff is however very small.

In [\ref{corsey1}] $e^+e^-$ interactions at a centre-of-mass energy
$\sqrt{s}=360$~GeV, slightly above the threshold for $t\bar t$
production, were considered. This is an ideal phenomenological
environment to test the impact of the implemented corrections, with
most of the radiation being associated to the top-decay stage.
Three-jet events were analysed and a remarkable impact of
matrix-element corrections was found when comparing different versions
of \herw \footnote{The latest public version \herw\ 5.9 had some
  errors in the treatment of top decays, which have been corrected in
  the intermediate version 6.0.  In order to estimate the impact of
  matrix-element corrections, we therefore compare the versions 6.1
  and 6.0}.  The results of \herw\ 6.1~[\ref{herwig61}], the new
version which includes also the improved treatment of top decays, were
also compared to the ones obtained by the exact ${\cal O}(\alpha_S)$
matrix-element calculation of the process $e^+e^-\to t\bar t\to
(bW^+)(bW^-) (g)$~[\ref{orr}].  While the authors of [\ref{orr}] had
found serious discrepancies when comparing the exact ${\cal
  O}(\alpha_s)$ results with the ones of \herw\ before matrix-element
corrections, good agreement was found in [\ref{corsey1}] after
matrix-element corrections in the region of large energies
and angles, where fixed-order calculations are reliable.

In [\ref{corsey1}] it was also shown that, although the fraction of
events generated in the dead zone varies from about 2\% to 4\% when
the cutoff changes from 5 GeV to 1 GeV, the dependence of
phenomenological distributions on its chosen value is pretty
negligible after one applies typical experimental cuts on the jet
transverse energy $E_T>10$~GeV and on the invariant opening angle
between jets $\Delta R>0.7$.  The value $E_{\mathrm{min}}=2$~GeV was
then chosen as the default value and this value will be kept throughout
this paper as well.  In the following sections, we shall consider
hadronic production of $t\bar t$ pairs and analyse the effect of  
matrix-element
corrections to top decays on jet observables and on the top mass
reconstruction at the Tevatron and at the
LHC.

\section{Jet activity in dilepton \boldmath$t\bar t$ events}
We start our analysis by considering inclusive jet observables. To
emphasize the effect of the matrix element corrections to top decays,
we confine ourselves to the case of leptonic decays of both $W$'s in
each event.  The most likely hard and well-separated jets are then
those from the $b$ and the $\bar b$. Extra radiation from either the
initial state (ISR) or the top decay may give rise to extra jets (final
state radiation from the produced $t\bar t$ pair tends to be small
because of the large top mass).
Interesting observables which may show the effects of the new \herw\ 
treatment of top decays are related to these extra jets, and in
particular to the one with the largest value of $E_T$, the `third jet'.

We cluster jets according to the inclusive version of the $k_T$
algorithm [\ref{kt}], setting a radius parameter $R=0.5$ at the
Tevatron and $R=1$ at the LHC\footnote{Such a choice is due to the
  relation between the $R$ parameter of the $k_T$ algorithm and the
  radius $R_{\mathrm{cone}}$ of the cone algorithm  
$R_{\mathrm{cone}}\approx
  0.75\times R$ [\ref{soper}] and to the fact that the
  Tevatron experimentalist run a pure cone algorithm with
  $R_{\mathrm{cone}}=0.4$.  For the LHC we shall nevertheless stick  
to
  the recommended value $R=1$.}.  We set a cutoff on the transverse
energy of the resolved jets $E_T>10$~GeV, and we study the following
inclusive distributions: transverse energy $E_T$ and rapidity $\eta_3$
of the third jet, the minimum invariant opening angle $\Delta R=
\sqrt{\Delta\phi^2+\Delta\eta^2}$ among the three hardest jets, the
threshold variable $d_3$ of the $k_T$ algorithm, according to which
all events are forced to be three-jet-like.  Finally, we consider the
number of jets $n_{\mathrm{jets}}$ that pass the 10 GeV cut in
transverse energy.

\begin{figure}
\centerline{\resizebox{0.49\textwidth}{!}{\includegraphics{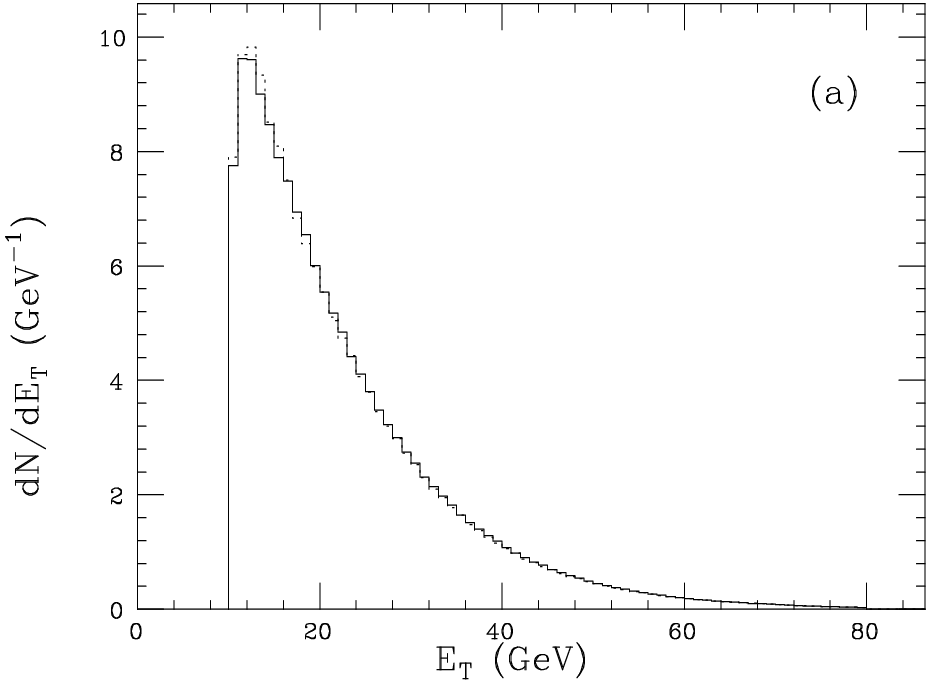}}%
\hfill%
\resizebox{0.49\textwidth}{!}{\includegraphics{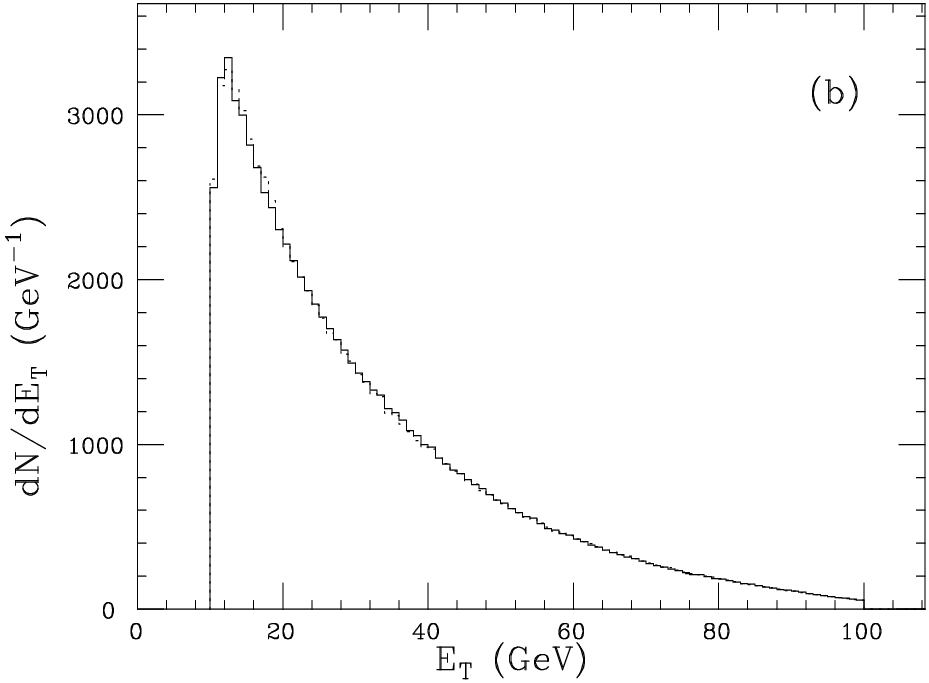}}}
  \caption{Transverse energy distribution of the third hardest jet at  
the 
    Tevatron (a) and at the LHC (b), according to \herw\ 6.1 (solid
    line) and 6.0 (dotted).}
\label{et}
\end{figure}

\begin{figure}
\centerline{\resizebox{0.49\textwidth}{!}{\includegraphics{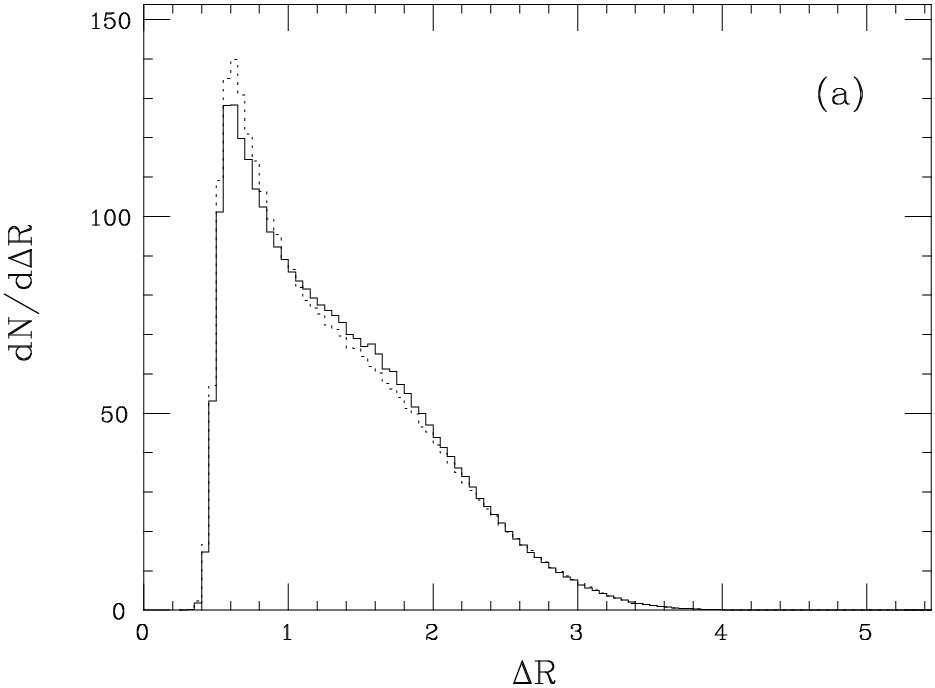}}%
\hfill%
\resizebox{0.49\textwidth}{!}{\includegraphics{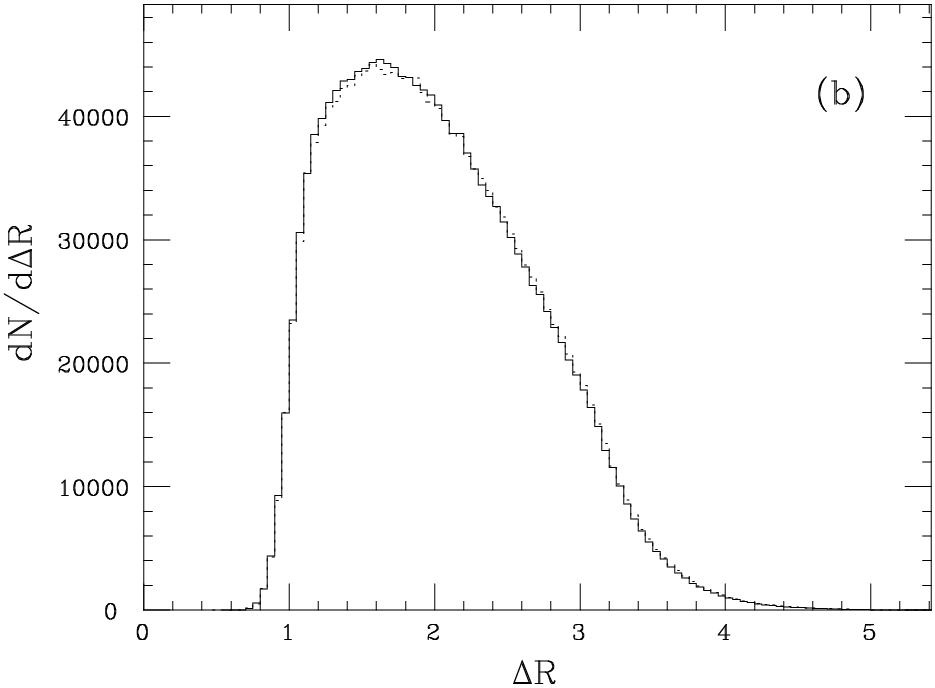}}}
  \caption{Distributions of the minimum invariant opening angle $\Delta  
R$ 
    among the three hardest jets at the Tevatron (a) and at the LHC
    (b), according to \herw\ 6.1 (solid line) and 6.0 (dotted).}
\label{deltar}
\end{figure}

\begin{figure}
\centerline{\resizebox{0.49\textwidth}{!}{\includegraphics{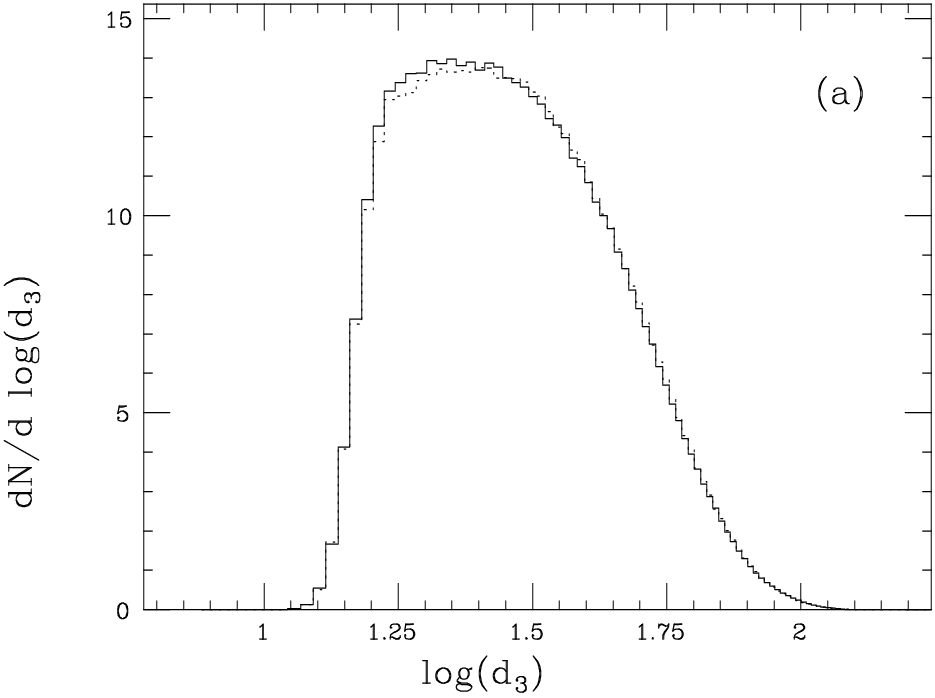}}%
\hfill%
\resizebox{0.49\textwidth}{!}{\includegraphics{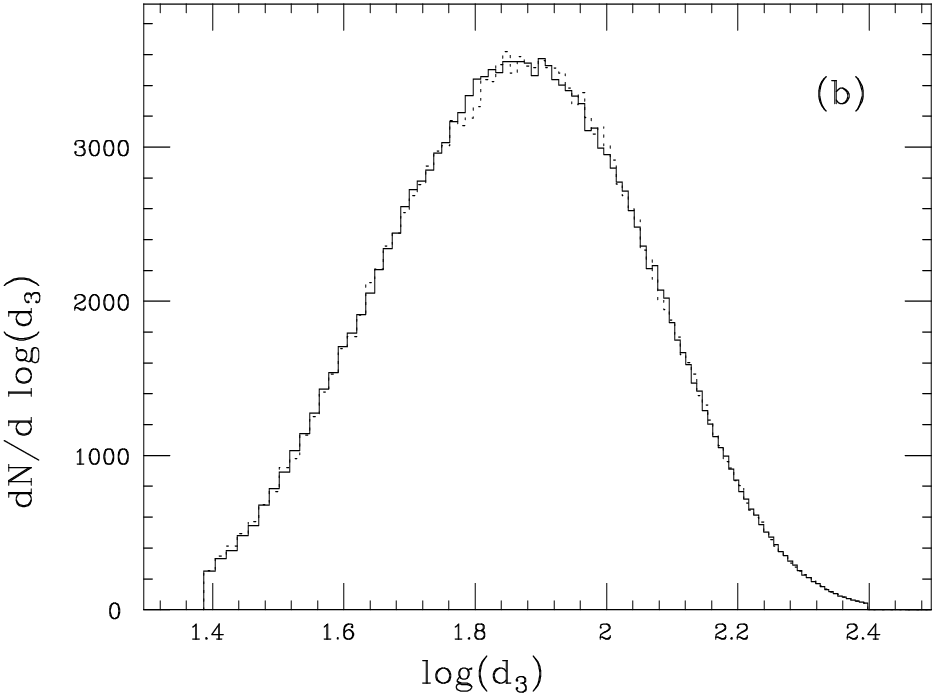}}}
  \caption{Distributions of the threshold variable $d_3$ for three-jet  
events
    at the Tevatron (a) and at the LHC (b), according to \herw\ 6.1
    (solid line) and 6.0 (dotted).}
\label{d3}
\end{figure}

\begin{figure}
\centerline{\resizebox{0.49\textwidth}{!}{\includegraphics{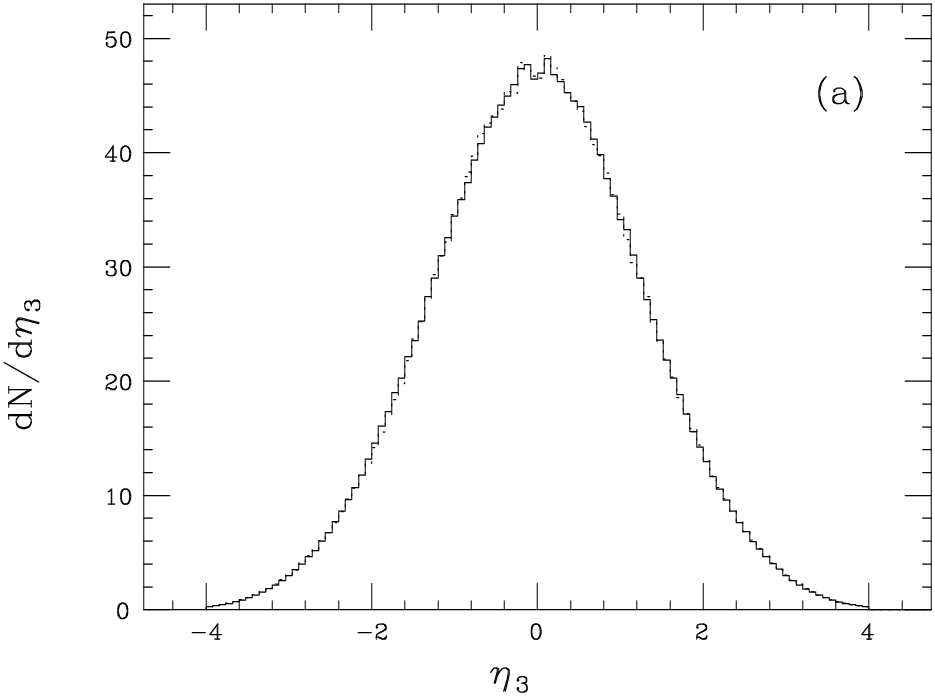}}%
\hfill%
\resizebox{0.49\textwidth}{!}{\includegraphics{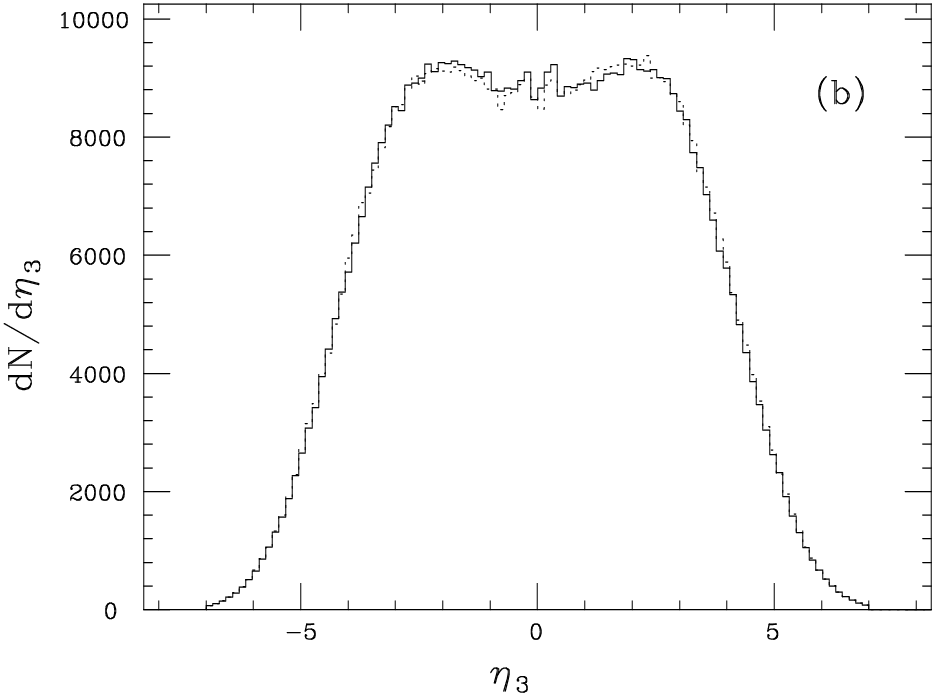}}}
\caption{Distributions of the rapidity of the third hardest jet
  at the Tevatron (a) and at the LHC (b), according to \herw\ 6.1
  (solid line) and 6.0 (dotted).}
\label{rap}
\end{figure}

\begin{figure}
\centerline{\resizebox{0.49\textwidth}{!}{\includegraphics{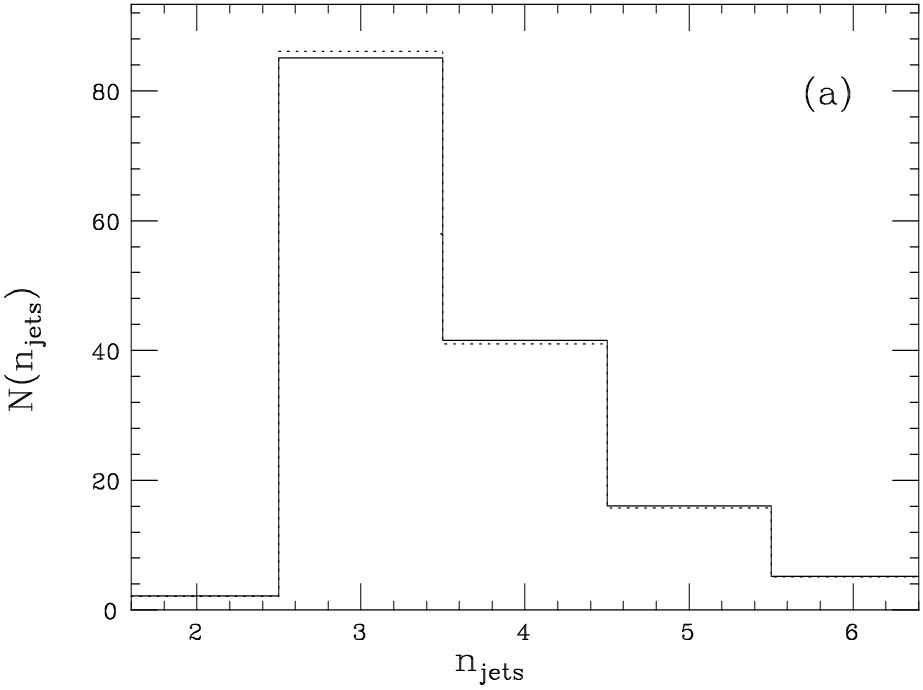}}%
\hfill%
\resizebox{0.49\textwidth}{!}{\includegraphics{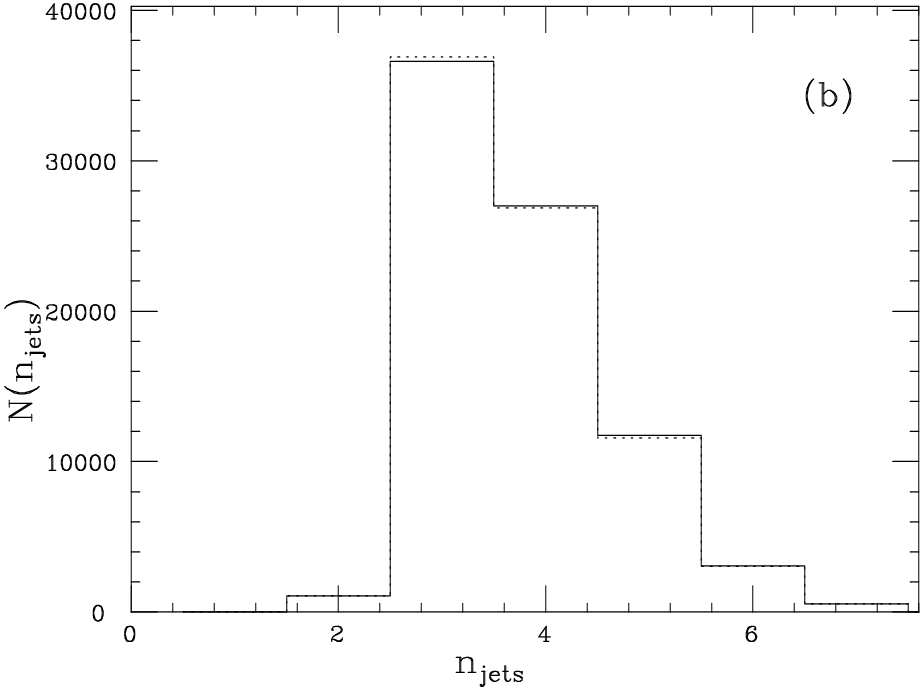}}}
  \caption{Number of jets passing the 10 GeV cut in transverse energy
    at the Tevatron (a) and at the LHC (b), according to \herw\ 6.1
    (solid line) and 6.0 (dotted).}
\label{njet}
\end{figure}

We start by comparing the Monte Carlo jet distributions before and
after matrix-element corrections. To make the comparison more
realistic, we normalize the plots to the expected integrated
luminosities (2~fb$^{-1}$ at $\sqrt{s}=2$~TeV for the Tevatron  
Run~II,
and 10~fb$^{-1}$ for 1 year of LHC low-luminosity
running~\footnote{Using total cross-sections of 7~pb and 830~pb for
  Tevatron and LHC~[\ref{bcmn}], and assuming a conservative 1\%
  overall efficiency and BR for the dilepton final states, this
  correponds to approximately 150 [\ref{cdf}] and 8$\times 10^{4}$
  [\ref{atlas}] events, respectively.}), and smear the contents of a
given bin with $N$ events according to a Gaussian distribution
with average $N$ and standard deviation $\sqrt{N}$.  Furthermore, to
partially account for detector effects we smear the value of the
reconstructed observables with a 10\% resolution. The final jet
distributions are plotted in Figs.~\ref{et}-\ref{njet}.
\footnote{With no smearing, the $\Delta R$ distributions at the
  Tevatron and at the LHC would have shown a sharp cutoff for $\Delta
  R=R$, as predicted by the $k_T$ algorithm we have been using, but
  nevertheless the 10\% smearing allows a small fraction of the jet
  events to have even $\Delta R<R$.}


All observables show very small changes due to the matrix-element
corrections. To quantify the residual differences, we performed a
Kolmogorov--Smirnov test on the distributions, and evaluated the number
of events necessary to establish the difference between the 6.0 and
6.1 distributions at 95\% CL. The results of this test are shown in
table \ref{kol}.  We see that at both machines a number of events
of the order of $\approx 10^3-10^4$ is sufficient to see a difference  
in
the shape of the two histograms. We therefore conclude that the effects  
of
matrix-element corrections to top decays are in principle detectable
at the LHC, where about $8\times 10^4$ events per year are expected in
the dilepton channel, while the foreseen statistics are too low at the
Tevatron.

\begin {table}
\begin{center}
\begin{tabular}{||l|r|r|r||}\hline
observable&Tevatron&LHC\\\hline
$E_T$&1000&1100\\\hline
$\log(d_3)$&3000&2700\\\hline
$\Delta R$&1100&4500\\\hline
$\eta_3$&5300&2500\\\hline
\end{tabular}
\end{center}
\caption{Number of $t\bar t$ events for which, according to the  
Kolmogorov 
  test, the jet distributions using \herw\ 6.0 and 6.1 are different
  at the 95\% confidence level.\label{kol}}
\end{table}
We conclude this Section by pointing out that the small differences
detected in the case of inclusive third-jet observables are largely a
consequence of the small fraction of events for which the
matrix-element corrections are applied. We evaluated in fact that for
the Tevatron energy only approximately 7\% of events that contain a
third jet
required the evaluation of the matrix-element corrections to the top
decays (the number is 6\% for the LHC).
We expect that larger effects and differences will appear once the
matrix-element corrections to the $t\bar t$ production process is
included. This work is under way, along similar lines to vector boson
production [\ref{corsey2}].

\section{Top mass reconstruction}
One observable for which the higher-order corrections described in
this paper may induce relevant changes is the invariant mass
distribution of $J/\psi+\ell$ pairs from $t\to b W$ decays, with the
$J/\psi$ coming from the decay of a $b$ hadron, and the isolated
lepton from the decay of the~$W$. This distribution was recently
suggested~[\ref{avto}] as a potential way of measuring the top mass
during the high-luminosity phase of the LHC with experimental
accuracies better than 1~GeV. The dominant source of experimental
uncertainty is the statistical one, due to the strong suppression from
the small branching ratios~\footnote{Ref.~[\ref{avto}] estimates a
  sample of approximately $10^3$ events in one year of high
  luminosity running at the LHC, ${\cal L} = 10^5$  
${\mathrm{pb}}^{-1}$,
  for a production cross section $\sigma_{\mathrm{LHC}} (t\bar
  t)=833$~pb and a total branching fraction $B=5.3\times 10^{-5}$,
  using the current expectations for tracking and reconstruction
  efficiencies.} . The shape of the $m_{J/\psi+\ell}$ distribution can
be compared with a template of shapes parametrised by the top mass,
and the value of $m_t$ can therefore be fitted. The Lorentz
invariance of the observable makes it completely independent of the
details of the $t\bar t$ production mechanism, and of the structure of
the ISR.  Given that the spectrum of the leptons from the $W$ decay is
known very well, the dominant theoretical uncertainty comes therefore
from the predictions for the spectrum of the $J/\psi$. This is obtained
from the convolution of the energy spectrum of $b$ hadrons in the top
decay, with the spectrum of $J/\psi$'s in the $b$-hadron decay. In  
principle
this second element can be measured with high accuracy over the next
few years at the $B$-factories~\footnote{Although small corrections
  are expected to be needed for this application, due to the different
  composition of $b$-hadrons in top decays relative to that in the
  decays of the $\Upsilon(4S)$.}.

We shall therefore concentrate here on the problem of the $b$-hadron
spectrum in top decays, and investigate how the top mass measurement
is affected by the matrix-element corrections to top decays in \herw.
For simplicity, we shall analyse the $m_{B\ell}$ spectra (instead of
the
$m_{J/\psi+\ell}$ ones) obtained by running \herw\ with and without
matrix-element corrections to top decays.

The Tevatron statistics will be too low to use this channel as a probe
of the top mass. Nevertheless we shall present results for the
Tevatron as well, to show that indeed the details of the production
mechanism (which is mainly $q\bar q\to t\bar t$ at the Tevatron and
$gg\to t\bar t$ at the LHC) have no impact on the top mass
determination. Since in $q\bar q$ annihilation the $t\bar t$ pair is
always produced in a colour octet state, while in $gg$ fusion it may
come either in a colour singlet or colour octet state, this comparison
will indicate that non-perturbative corrections
to the $m_{B\ell}$ mass spectrum are very weakly dependent on the
details of the colour-neutralization model in the MC.

\begin{figure}
\centerline{\resizebox{0.49\textwidth}{!}{\includegraphics{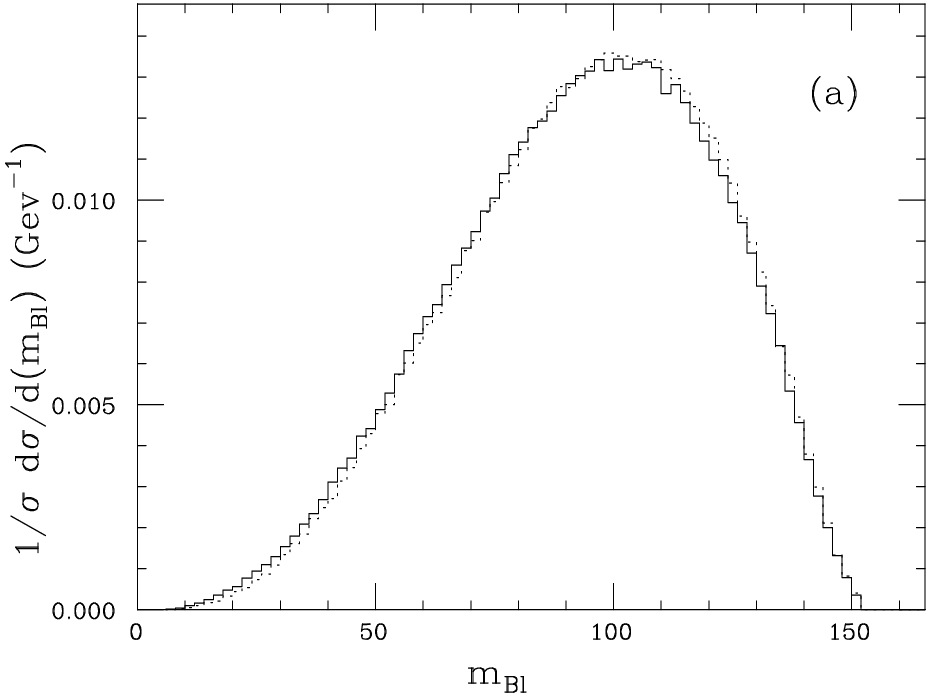}}%
\hfill%
\resizebox{0.49\textwidth}{!}{\includegraphics{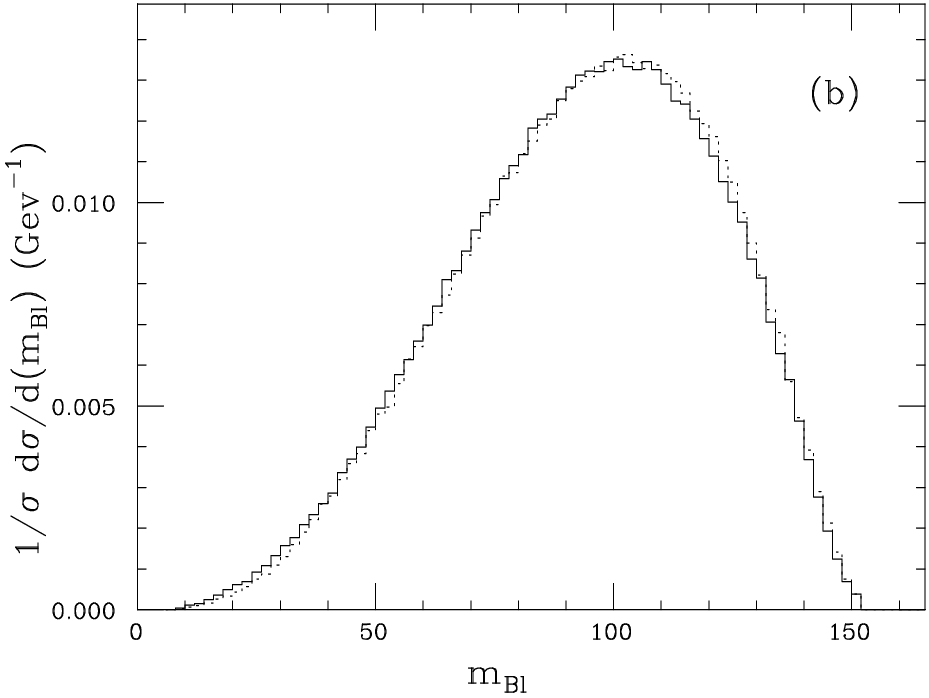}}}
  \caption{Invariant mass of the $B$-lepton system at the Tevatron (a)
    and at the LHC (b) for $m_t=175$~GeV, according to \herw\ 6.0
    (dotted) and 6.1 (solid).}
  \label{mass}
\end{figure}

\begin{figure}
\centerline{\resizebox{0.65\textwidth}{!}{\includegraphics{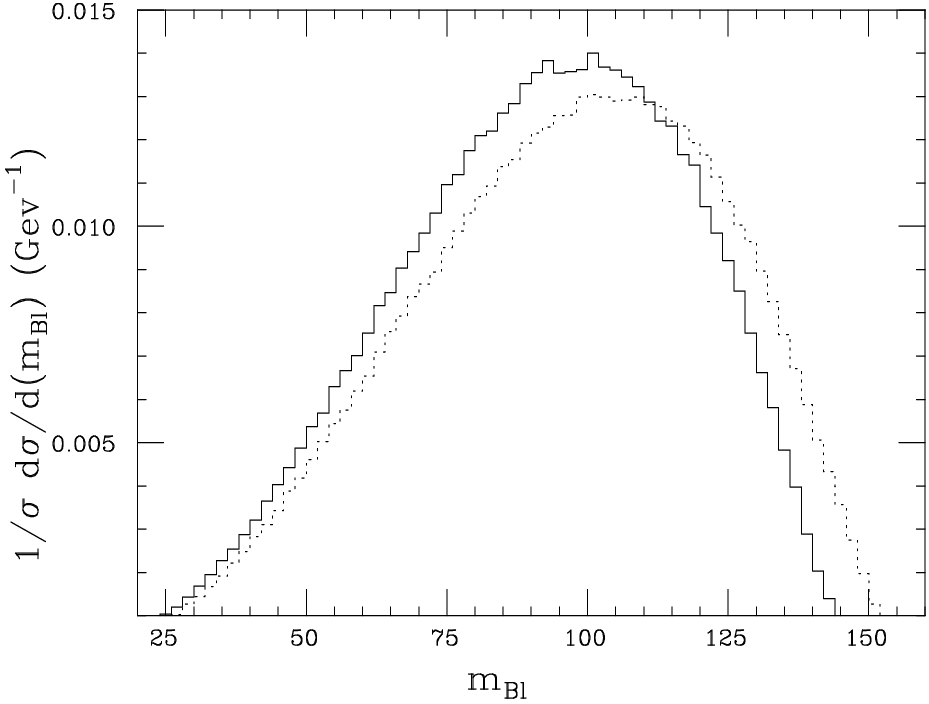}}}
  \caption{Invariant mass distributions according to \herw\ 6.1 at the  
LHC
 for $m_t=171$~GeV (solid) and 179 GeV (dotted).}
  \label{mass19}
\end{figure}

We generate $t\bar t$ samples using \herw\ 6.0 and 6.1 and plot
the $m_{B\ell}$ spectra for different values of $m_t$.
We then evaluate the average value and the standard
deviation of our distributions, the differences in the average
values $\langle m_{B\ell}\rangle$ and the corresponding statistical  
errors.  In
Fig.~\ref{mass} the $m_{B\ell}$ spectra are plotted for  
$m_t=175$~GeV,
at the Tevatron and at the LHC, before and after matrix-element  
corrections,
while in Fig.~\ref{mass19} one can find the distributions at the LHC, 
according to \herw\ 6.1, for $m_t=171$~GeV and $m_t=179$~GeV.
In tables~\ref{masstev} and \ref{masslhc} we
summarize the results of our statistical analysis.
\begin{table}
\begin{center}
\begin{tabular}{||l|r|r|r|r|r|r||}\hline
\ \ \  $m_t$\ \ \
&$\langle m_{B\ell}\rangle^{6.1}$&$\sigma(6.1)$&$\langle  
m_{B\ell}\rangle^{6.0}$
&$\sigma(6.0)$&$\langle m_{B\ell}\rangle^{6.0}-\langle
m_{B\ell}\rangle^{6.1}$\\  \hline
171 GeV&91.18 GeV&26.51 GeV&92.06 GeV&26.17 GeV&$(0.873\pm 0.037)$  
GeV\\\hline
173 GeV&92.31 GeV&26.90 GeV&93.22 GeV&26.58 GeV&$(0.912\pm 0.038)$  
GeV\\\hline
175 GeV&93.41 GeV&27.29 GeV&94.38 GeV&26.94 GeV&$(0.972\pm 0.038)$  
GeV\\\hline
177 GeV&94.65 GeV&27.73 GeV&95.45 GeV&27.33 GeV&$(0.801\pm 0.039)$  
GeV\\\hline
179 GeV&95.64 GeV&28.00 GeV&96.63 GeV&27.60 GeV&$(0.984\pm 0.039)$  
GeV\\\hline
\end{tabular}
\end{center}
\caption{Results at the Tevatron
 for different values of $m_t$.\label{masstev}}
\end{table}%
\begin {table}
\begin{center}
\begin{tabular}{||l|r|r|r|r|r|r||}\hline
\ \ \ $m_t$\ \ \
&$\langle m_{B\ell}\rangle^{6.1}$&$\sigma(6.1)$&
$\langle m_{B\ell}\rangle^{6.0}$&
$\sigma(6.1)$&$\langle m_{Bl}\rangle^{6.0}-\langle  
m_{Bl}\rangle^{6.1}$\\ \hline
171 GeV&91.13 GeV&26.57 GeV&92.02 GeV&26.24 GeV&$(0.891\pm 0.038)$  
GeV\\\hline
173 GeV&92.42 GeV&26.90 GeV&93.26 GeV&26.59 GeV&$(0.844\pm 0.038)$  
GeV\\\hline
175 GeV&93.54 GeV&27.29 GeV&94.38 GeV&27.02 GeV&$(0.843\pm 0.039)$  
GeV\\\hline
177 GeV&94.61 GeV&27.66 GeV&95.46 GeV&27.33 GeV&$(0.855\pm 0.039)$  
GeV\\\hline
179 GeV&95.72 GeV&28.04 GeV&96.51 GeV&27.67 GeV&$(0.792\pm 0.040)$  
GeV\\\hline
\end{tabular}
\end{center}
\caption{As in table 1, but for the LHC.\label{masslhc}}
\end{table}%
We observe a systematic shift of about $800-900$~MeV towards lower
values of $\langle m_{B\ell}\rangle$ after matrix-element corrections  
to
top decays. Furthermore, the results at the Tevatron are the same as
the ones at the LHC, to within 150~MeV. 
Using the Kolmogorov--Smirnov test, we find that about  
$N_{\mathrm{eff}}\simeq
6000$ reconstructed final states are sufficient to verify that the
shapes of the 6.1 and the 6.0 distributions are different at the
confidence level of 95\%; one year of high-luminosity run would  
nevertheless 
allow one to distinguish the two distributions at 70\% confidence  
level.
\begin {table}
\begin{tabular}{||l|r|r|r|r|r|r||}\hline
$m_t$&$\langle m_{B\ell}\rangle^{6.1}$&$\sigma (6.1)$&
$\langle m_{B\ell}\rangle^{6.0}$&$\sigma (6.0)$&
$\langle m_{B\ell}\rangle^{6.0}-\langle m_{B\ell}\rangle^{6.1}$\\
\hline
171 GeV&95.98 GeV&22.22 GeV&96.43 GeV&22.22 GeV&$(0.476\pm 0.036)$  
GeV\\\hline
173 GeV&97.03 GeV&22.64 GeV&97.51 GeV&22.68 GeV&$(0.511\pm 0.034)$  
GeV\\\hline
175 GeV&98.08 GeV&23.15 GeV&98.59 GeV&23.11 GeV&$(0.514\pm 0.035)$  
GeV\\\hline
177 GeV&99.23 GeV&23.58 GeV&99.60 GeV&23.53 GeV&$(0.396\pm 0.035)$  
GeV\\\hline
179 GeV&100.13 GeV&23.93 GeV&100.63 GeV&23.90 GeV&$(0.560\pm 0.036)$  
GeV\\\hline
\end{tabular}
\caption{Results for the invariant mass $m_{B\ell}$ at the Tevatron for  
different values of $m_t$, once we select a sample with
  $m_{B\ell}>50$~GeV.\label{tev50}}
\end{table}
\begin {table}
\begin{tabular}{||l|r|r|r|r|r|r||}\hline
$m_t$&$\langle m_{B\ell}\rangle^{6.1}$&$\sigma (6.1)$&
$\langle m_{B\ell}\rangle^{6.0}$&$\sigma (6.0)$&
$\langle m_{B\ell}^{6.0}\rangle-\langle m_{B\ell}^{6.1}\rangle$\\  
\hline
171 GeV&95.97 GeV&22.24 GeV&96.45 GeV&22.26 GeV&$(0.479\pm 0.036)$  
GeV\\\hline
173 GeV&97.09 GeV&22.69 GeV&97.56 GeV&22.68 GeV&$(0.479\pm 0.034)$  
GeV\\\hline
175 GeV&98.14 GeV&23.12 GeV&98.64 GeV&23.15 GeV&$(0.510\pm 0.035)$  
GeV\\\hline
177 GeV&99.16 GeV&23.54 GeV&99.62 GeV&23.52 GeV&$(0.466\pm 0.035)$  
GeV\\\hline
179 GeV&100.20 GeV&23.96 GeV&100.62 GeV&23.90 GeV&$(0.427\pm 0.036)$  
GeV\\\hline
\end{tabular}
\caption{Results for the invariant mass $m_{B\ell}$ at the LHC for  
different 
values of $m_t$, once we select a sample with $m_{B\ell}>50$~GeV.
\label{lhc50}}
\end{table}
\par

\begin{figure}
\centerline{\resizebox{0.65\textwidth}{!}{\includegraphics{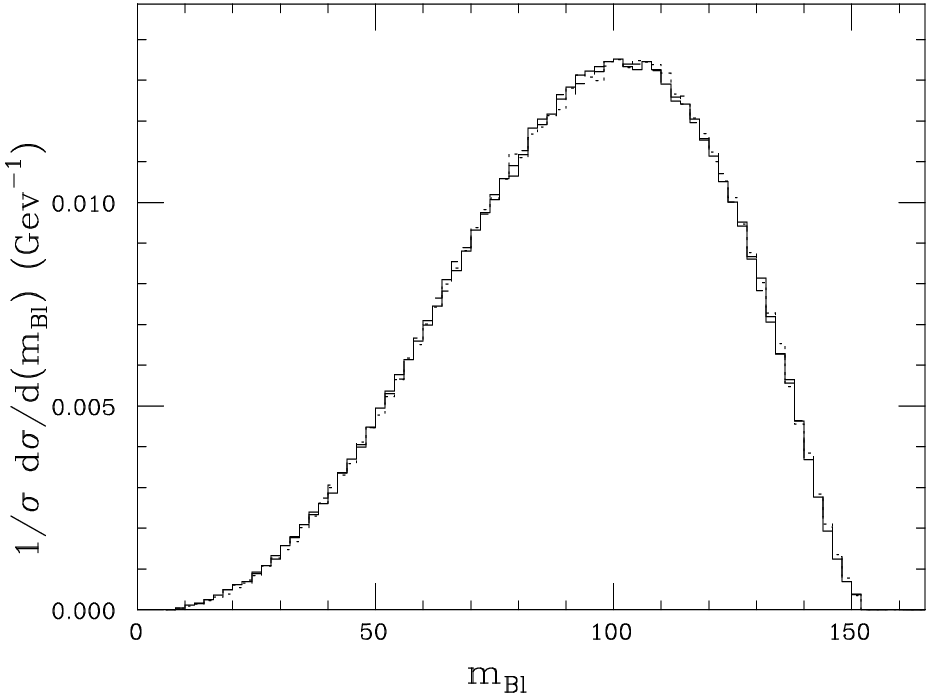}}}
  \caption{Invariant mass distributions according to \herw\ 6.1 for
    $m_t=175$~GeV and a cutoff on the backward gluon energy equal to
      1 GeV (dotted line), 2 GeV (solid) and 5 GeV (dashed).}
  \label{cutoff}
\end{figure}

We studied the dependence of our results on the chosen infrared cutoff
for the energy of the gluons emitted in the dead zone, which was
discussed in Section 2. We find a negligible variation: at the LHC,
and for $m_t=175$~GeV, we find $\langle m_{B\ell}\rangle=93.78,\,
93.54$ and 93.47~GeV for $E^{\rm cutoff}=1,\, 2$ and 5~GeV,  
respectively. 
The distributions in
Fig.~\ref{cutoff}, obtained for $m_t=175$~GeV and different values of
$E^{\rm cutoff}$, are essentially identical.

If we set a cut $m_{B\ell}>50$~GeV on the invariant mass, to reduce
the sensitivy to low-mass tails possibly affected by backgrounds, we
find the results summarized in tables~\ref{tev50} and \ref{lhc50} for
the Tevatron and the LHC respectively.  Once we cut off part of the
spectrum, it is to be expected that we find higher values for $\langle
m_{B\ell}\rangle$ and lower values for the differences $\langle
m_{B\ell}\rangle^{6.0}-\langle m_{B\ell}\rangle^{6.1}$, which are  
indeed now of
about $400-500$~MeV. Once again the shifts at the
Tevatron and at the LHC are of similar size, within 150~MeV. 

\begin{figure}
\centerline{\resizebox{0.65\textwidth}{!}{\includegraphics{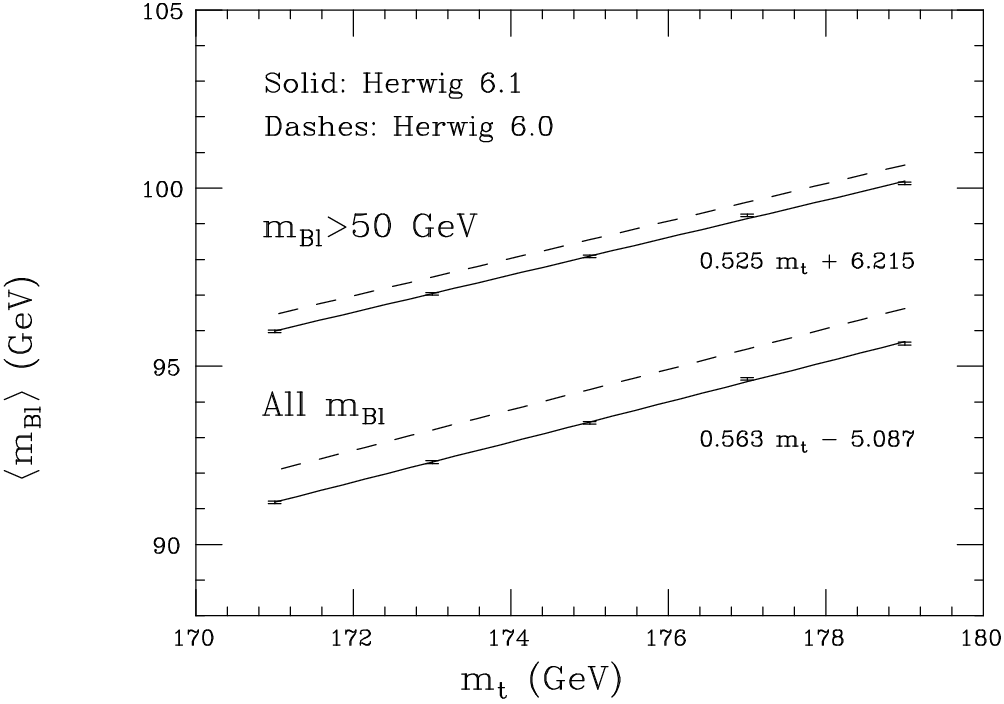}}}
  \caption{Results for the average invariant mass $\langle
    m_{B\ell}\rangle$ as a function of $m_t$ at the Tevatron after a  
fit
    into a straight line.  The solid and dashed lines refer to \herw\ 
    6.1 and \herw\ 6.0 respectively.}
  \label{fittev}
\end{figure}

\begin{figure}
\centerline{\resizebox{0.65\textwidth}{!}{\includegraphics{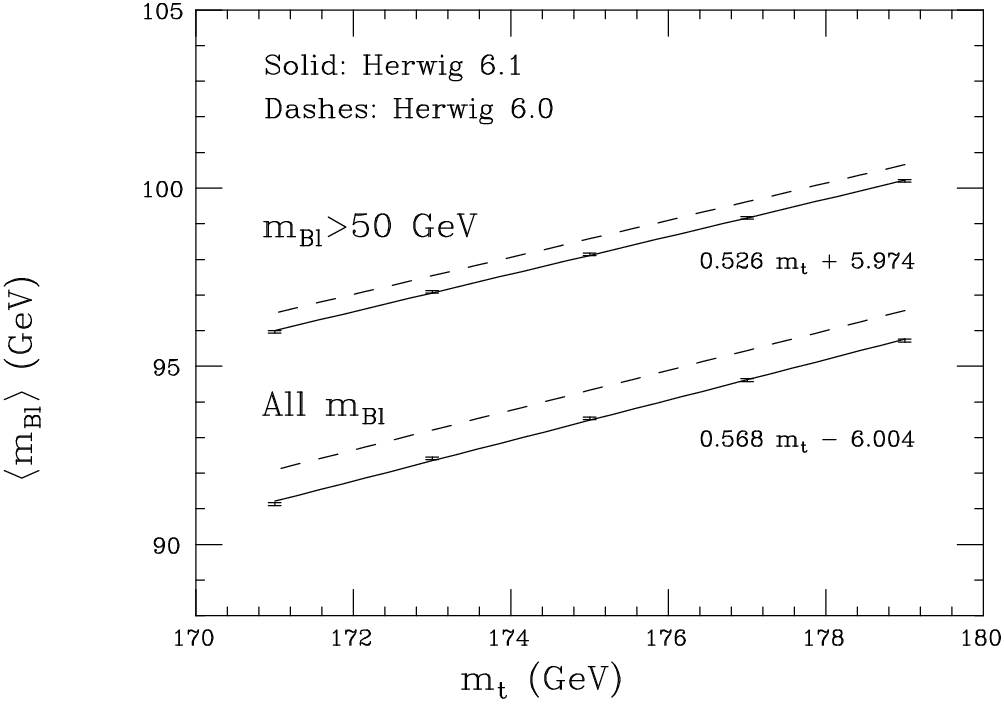}}}
  \caption{Results for the average invariant mass $\langle
    m_{Bl\ell}\rangle$ as a function of $m_t$ at the LHC after a fit  
into
    a straight line.  The solid and dashed lines refer to \herw\ 6.1
    and \herw\ 6.0 respectively.}
  \label{fitlhc}
\end{figure}

In order to evaluate the impact of the found discrepancies on the top
mass, we perform a linear fit of the $\langle m_{B\ell}\rangle$
distribution as a function of $m_t$, by means of the least square
method.  We find, after considering all the $m_{B\ell}$ values:
\begin{eqnarray} 6.1\ :\;
\langle m_{B\ell}\rangle&=&0.563\ m_t-5.087\ {\mathrm{GeV}}\ ,\
\epsilon({\mathrm{GeV}})=0.046\ 
{\mathrm{(Tevatron)}}\  ;\\
6.0\ :\;
\langle m_{B\ell}\rangle&=&0.568\ m_t-5.139\ {\mathrm{GeV}}\ ,\
\epsilon({\mathrm{GeV}})=0.023 \ 
{\mathrm{(Tevatron)}}\ ;\\
6.1\ :\;
\langle m_{B\ell}\rangle&=&0.568\ m_t-\ 6.004\ {\mathrm{GeV}}\ ,\
\epsilon({\mathrm{GeV}})=0.057\ 
{\mathrm{(LHC)}}\ ;\\
6.0\ :\;
\langle m_{B\ell}\rangle&=&0.559\ m_t-3.499\ {\mathrm{GeV}}\ ,\
\epsilon({\mathrm{GeV}})=0.052\ 
{\mathrm{(LHC)}}\  ;
\end{eqnarray}
where $\epsilon$ is the mean square deviation in the fit. We see that
the linear fit is very good, and well within the required accuracy.
The obtained fits are plotted in Figs.~\ref{fittev} and \ref{fitlhc}
at the Tevatron and at the LHC respectively. The error bars correspond
to the statistical errors found on $\langle
m_{B\ell}^{6.0}\rangle-\langle m_{B\ell}^{6.1}\rangle$, which, as can  
be
seen from the figures, are significantly lower than the difference due
to the implementation of matrix-element corrections to top decays.
This means that the impact on the top mass is a physical effect and
not just the result of statistical fluctuations.

Using these fits, and inverting the relation between $\langle
m_{B\ell}\rangle$ and $m_t$, we find that for a given value of $\langle
m_{B\ell}\rangle$ consistent with the range $171\lsim m_t \lsim 179$,  
the
values of $m_t$ extracted using the two versions of \herw\ differ by
1.5~GeV, both at the Tevatron and at the LHC. This is a rather large
value, competitive with the expected systematic error at the LHC,
indicating that such corrections are relevant, and must be applied.

After setting a cut of 50 GeV on the $m_{Bl}$ spectra, we obtain the
following fits:
\begin{eqnarray} 6.1\ :\;
\langle m_{B\ell}\rangle&=&0.525\ m_t+6.125\ {\mathrm{GeV}}\ ,\ 
\epsilon({\mathrm{GeV}})=0.049\ 
{\mathrm{(Tevatron)}}\  ;\\
6.0\ :\;
\langle m_{B\ell}\rangle&=&0.524\ m_t+6.765\ {\mathrm{GeV}}\ ,\ 
\epsilon({\mathrm{GeV}})=0.022\ 
{\mathrm{(Tevatron)}}\ ;\\
6.1\ :\;
\langle m_{B\ell}\rangle&=&0.526\ m_t+5.974\ {\mathrm{GeV}}\ ,\
\epsilon({\mathrm{GeV}})=0.026\ 
{\mathrm{(LHC)}}\ ;\\
6.0\ :\;
\langle m_{B\ell}\rangle&=&0.520\ m_t+7.578\ {\mathrm{GeV}}\ ,\
\epsilon({\mathrm{GeV}})=0.040\ 
{\mathrm{(LHC)}}\  .
\end{eqnarray}
The differences in the top mass extraction between the 6.0 and 6.1
versions from a given value of $\langle m_{B\ell}\rangle$ are now
reduced to 1~GeV, smaller than for the
fully inclusive distribution, but still significant relative to the
overall accuracy goal of 1~GeV.

\par

To conclude our study, we wish to investigate the dependence of our
results on the hadronisation model used by \herw. We start by studying
the invariant mass distribution of the lepton with the $b$ quark, as
opposed to the $b$-hadron. We do this by considering the \herw\ final
state at the end of the perturbative evolution, just before the
non-perturbative gluon-splitting phase which precedes the formation of
the colour-singlet clusters, and the eventual hadronisation. 

\begin{figure}
\centerline{\resizebox{0.49\textwidth}{!}{\includegraphics{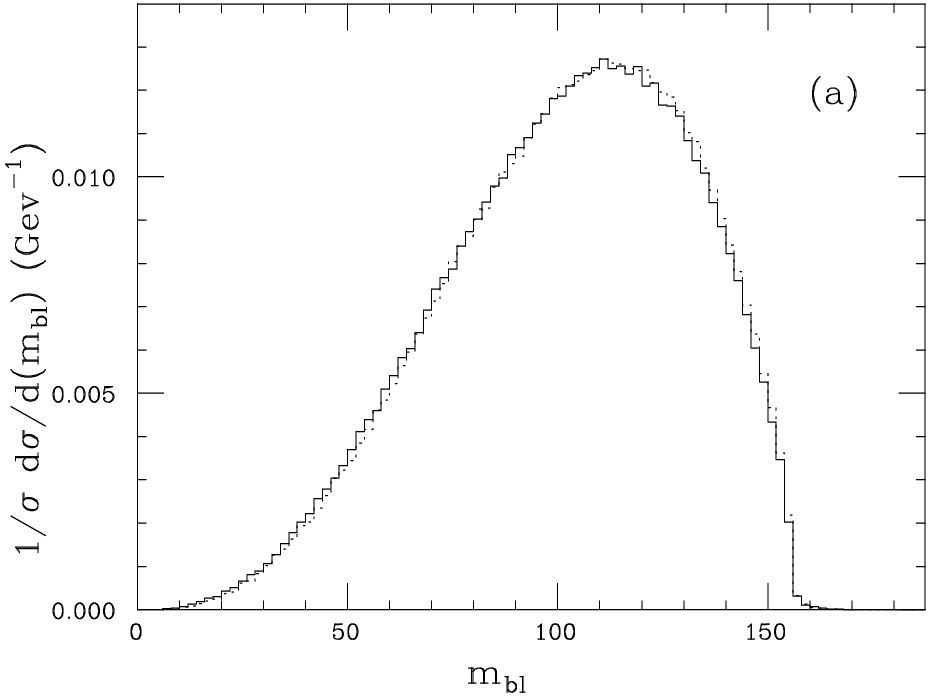}}%
\hfill%
\resizebox{0.49\textwidth}{!}{\includegraphics{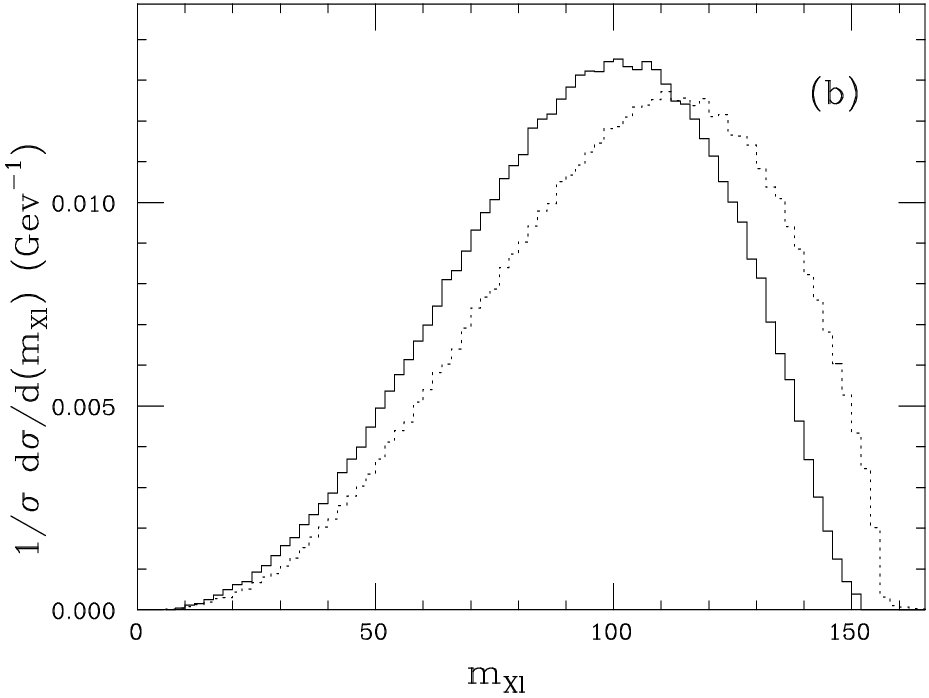}}}
  \caption{Distributions of the invariant mass of the $b$-lepton 
    system for $m_t=175$~GeV at the LHC according to \herw\ 6.1  
(solid
    line) and 6.0 (dotted) (a) and according to 6.1, but at  
hadron-level
($X=B$, solid) and parton level ($X=b$, dotted) (b).}
  \label{masspl}
\end{figure}

In Fig.~\ref{masspl} we plot the 6.0 and 6.1 distributions of the 
invariant mass of the $b$-lepton system for
$m_t=175$~GeV and the 6.1 ones at hadron- and parton-level.
The results at parton level for different values of the top mass can be
found in table~\ref{tabmbl}.  We find that, at fixed $m_t$, the
average values of the parton-level invariant masses 
$\langle m_{b\ell}\rangle$ are larger than
the ones after the hadronisation of the $b$ quark, with differences 
between the 6.0 and the 6.1 version of the order about
$600-700$~MeV. 

After a linear fit, we find the following relations:
\begin{eqnarray} 
6.1\ :\;
\langle m_{b\ell}\rangle &=&0.640\ m_t-10.256\ {\mathrm{GeV}}\ ,\
\epsilon({\mathrm{GeV}})=0.050\ . \\
6.0\ :\;
\langle m_{b\ell}\rangle &=&0.646\ m_t-10.528\ {\mathrm{GeV}}\ ,\
\epsilon({\mathrm{GeV}})=0.033
\end{eqnarray}
Given the values of the slopes, the shift in the extracted top mass  
between 
the two versions is now of about 
1.0~GeV, compared to the 1.5~GeV after hadronisation.

As a whole, the non-perturbative contribution to
the shape of the $b$-hadron spectra is therefore pretty important,
as is already known in the case of $Z^0$ decays.
Confidence in the accurate description of the non-perturbative
phase should be gained from the study of the $B$-fragmentation function
in $Z^0$ decays. Our study of the impact of matrix element corrections
at the Tevatron and at the LHC suggests that non-perturbative
corrections do not depend significantly on the production
mechanism. This we also checked by performing a similar study in the
case of $e^+e^-$ production. We therefore expect that once a
tuning of the non-perturbative $b$ fragmentation function in \herw\ is
achieved, using for example the latest high-precision SLD
results~[\ref{SLD}], the results can be extended to the study of the 
$B\ell$ mass spectrum in top decays. Given the size of the mass shift
induced by hadronisation (of the order of 8~GeV, as found by comparing
Table~\ref{tabmbl} and \ref{masslhc}), a control over the
fragmentation function at a level better than 5\% will have to be
achieved in order to maintain this contribution to the theoretical
systematics well below ${\cal O}$(1~GeV). This should be possible,  
since
the current size of the experimental uncertainty on $\langle x_B
\rangle$ is at the level of $\sim 1$\%~[\ref{SLD}].

\begin {table}
\begin{tabular}{||l|r|r|r|r|r|r||}\hline
$m_t$&$\langle m_{b\ell}\rangle^{6.1}$&$\sigma (6.1)$&$\langle
m_{b\ell}\rangle^{6.0}$&$\sigma (6.0)$& $\langle
m_{b\ell}\rangle^{6.0}-\langle m_{b\ell}\rangle^{6.1}$\\ \hline 
171 GeV&99.19 GeV&28.42 GeV&99.81 GeV&28.20 GeV&$(0.627\pm 0.040)$  
GeV\\\hline
173 GeV&100.47 GeV&28.84 GeV&101.17 GeV&28.43 GeV&$(0.701\pm 0.041)$  
GeV\\\hline
175 GeV&101.76 GeV&29.24 GeV&102.48 GeV&29.01 GeV&$(0.718\pm 0.041)$  
GeV\\\hline
177 GeV&102.93 GeV&29.67 GeV&103.72 GeV&29.43 GeV&$(0.791\pm 0.042)$  
GeV\\\hline
179 GeV&104.36 GeV&30.11 GeV&104.99 GeV&29.87 GeV&$(0.628\pm 0.043)$  
GeV\\\hline
\end{tabular}
\caption{Results for the invariant mass $m_{b\ell}$ at the LHC for
  different values of $m_t$.
\label{tabmbl}}
\end{table}
\section{Conclusions}
We have studied $t\bar t$ events in the dilepton channel at the
Tevatron and at the LHC using the new version of the \herw\ Monte
Carlo event generator, which includes matrix-element corrections to
the description of top decays.  We considered observables involving
the third hardest jet in transverse energy, to enhance possible
effects of the implemented corrections.

We have found that the distributions obtained before and after
matrix-element corrections are rather similar. A Kolmogorov test,
which compares the shapes of two distributions, allows however to
detect differences at the  95\%CL with the large statistics available 
at the LHC. The statistics of the Run~II at the Tevatron are however
not sufficient.

We have also investigated the reconstruction of the top mass by
looking at final states with leptons and a $J/\psi$. These final
states are an excellent candidate for an experimental determination of
the top mass with systematic errors in the range of 1~GeV. While
the main production mechanisms of top quarks are
different at the Tevatron and at the LHC, we found
equivalent results in the two cases, indicating 
that this method of reconstruction of 
the top mass is not sensitive to the details of the
colour-neutralisation model.

We considered the spectra of the invariant mass $m_{B\ell}$, where the
$B$ meson comes from the hadronisation of the $b$ quark produced in
$t\to bW$ and $\ell$ is the charged lepton from the decay $W\to \ell\nu$,
for different values of $m_t$ and obtained that the implementation of
matrix-element corrections to top decays results in a shift of about
1.5 GeV on the top mass if one is able to reconstruct the whole
$m_{B\ell}$ spectrum and of about 1 GeV after setting the cut
$m_{B\ell}>$~50 GeV.  The shifts we found are physical effects
related to the inclusion of hard and large-angle gluon radiation in
the Monte Carlo shower, since they are much larger than the
statistical errors on them.  Analyses at the parton level have shown an
impact of a similar magnitude.

It will be now very interesting to compare the new \herw\ results for
the invariant-mass distributions with the ones obtained after
performing a next-to-leading order calculation of the process $t\to
bWg$, and convoluting the result with the fragmentation function for
the hadronisation of the $b$ quark into a $B$ meson taken from 
LEP and SLD data.  In order to perform such a comparison in detail, the  
\herw\
cluster model used to simulate the hadronisation process will have to
be tuned to fit that data.

Furthermore, although in this paper we have concentrated our analysis
on the top mass reconstruction in final states with leptons and
$J/\psi$, it will be worthwhile redoing the Tevatron analysis to
determine $m_t$ using the \herw\ parton shower model, provided with
matrix-element corrections.

We finally recall that, though we feel safely confident that the new
version of \herw\ will be a trustworthy event generator for the
purpose top decays, top production is still performed in the
soft/collinear approximation, with dead zones in the phase space which
are needed to be filled.  Matrix-element corrections to top production
are in progress and may have an impact on jet observables and on the
top mass measurement at the Tevatron as well.

\section*{Acknowledgements}
We wish to acknowledge Bryan Webber for useful discussions on
these and related topics. We thank A. Kharchilava for several
discussions on the measurement of the top mass using the $\psi\ell$
final states.
G.C. is also grateful to the R.A.L. Theory
Group and to the CERN Theory Division for hospitality during some
of this work.

\section*{References}
\begin{enumerate}
\addtolength{\itemsep}{-0.5ex}
\item\label{frixione}
   S. Frixione, M.L. Mangano, P. Nason and G. Ridolfi, {\it Heavy
    Quark Production}, hep-ph/9702287, in {\it Heavy Flavours II},
  ed.\ A.J. Buras and M. Lindner, World Scientific, p.609; \\
S.~Frixione, M.~L.~Mangano, P.~Nason and G.~Ridolfi,
Nucl.\ Phys.\  {\bf B431} (1994) 453.
\item\label{Beneke:2000hk}
M.~Beneke {\it et al.}, hep-ph/0003033,
to appear in Proceedings of 1999 CERN Workshop on
Standard Model Physics (and more) at the LHC, G. Altarelli and
M.L. Mangano eds.
\item\label{Accomando:1998wt}
E.~Accomando {\it et al.}  [ECFA/DESY LC Physics Working Group
                  Collaboration],
Phys.\ Rept.\  {\bf 299}, 1 (1998)
[hep-ph/9705442].
\item\label{Mangano:1992jk}
M.~L.~Mangano, P.~Nason and G.~Ridolfi,
Nucl.\ Phys.\  {\bf B373}, 295 (1992); \\
S.~Frixione, M.~L.~Mangano, P.~Nason and G.~Ridolfi,
Phys.\ Lett.\  {\bf B351}, 555 (1995)
[hep-ph/9503213].
\item\label{Cacciari:1998it}
M.~Cacciari, M.~Greco and P.~Nason,
JHEP {\bf 9805}, 007 (1998)
[hep-ph/9803400]; \\
F.~I.~Olness, R.~J.~Scalise and W.~Tung,
Phys.\ Rev.\  {\bf D59}, 014506 (1999)
[hep-ph/9712494].
\item\label{herwig} 
 G. Marchesini and B.R. Webber, Nucl.\ Phys.\ B310 (1988) 461.
 G. Marchesini et al.,\ Comput.\  Phys.\  Commun.\  67 (1992) 465.
\item\label{pythia}
 T. Sj\"ostrand, Comp.\ Phys.\ Comm.\ 46 (1987) 367.
\item\label{Paige:1981fb}
F.~E.~Paige and S.~D.~Protopopescu,
BNL-29777;
H.~Baer, F.~E.~Paige, S.~D.~Protopopescu and X.~Tata,
\isajet~7.48, 
hep-ph/0001086.
\item\label{sey1}
 M.H. Seymour, Comput.\ Phys.\ Commun.\ 90 (1995) 95.
\item\label{sey2}
 M.H. Seymour, Z.\ Phys.\ C56 (1992) 161.
\item\label{sey3}
 M.H. Seymour, {\it Matrix Element Corrections to Parton Shower
Simulation of Deep Inelastic Scattering}, contributed to 27th
International Conference on High Energy Physics (ICHEP), Glasgow,
1994, Lund preprint LU-TP-94-12, unpublished.
\item\label{corsey1} 
 G. Corcella and M.H. Seymour, Phys.\ Lett.\ B442 (1998) 417.
\item\label{corsey2} 
 G. Corcella and M.H. Seymour, Nucl.\ Phys.\ B565 (2000) 227.
\item\label{mepapers}
  J. Andr\'e and T. Sj\"ostrand, Phys.\ Lett.\ B442 (1998) 417\\
  G. Miu and T. Sj\"ostrand, Phys.\ Lett.\ B449 (1999) 313. 
\item\label{avto}
 A. Kharchilava, Phys.\ Lett.\ B476 (2000) 73.
\item\label{corcella} 
  G. Corcella, hep-ph/9911447, {\it On the Top
    Mass Reconstruction Using Leptons}, contributed to the U.K.
  Phenomenology Workshop on Collider Physics, Durham, U. K., 19-24
  September 1999.
\item\label{marweb1}
G.Marchesini and B.R. Webber, Nucl. Phys. B330 (1990) 261
\item\label{marweb2}
 G. Marchesini and B.R. Webber, Nucl.\  Phys.\  B238 (1984) 1. 
\item\label{herwig61}
G. Corcella, I.G. Knowles, G. Marchesini, S. Moretti, K. Odagiri,
P. Richardson, M.H. Seymour, B.R. Webber, hep-ph/9912396.
\item\label{orr}
  L.H. Orr, T. Stelzer and W.J. Stirling, Phys.\ Rev.\ D56 (1997) 446
\item\label{kt} 
  S. Catani, Yu.L. Dokshitzer, M.H. Seymour and B.R.
  Webber,  Nucl.\ Phys.\ B406 (1993) 187\\
  M.H. Seymour, Nucl.\ Phys.\ B421 (1994) 545.
\item\label{soper} 
S.D. Ellis and D.E. Soper, Phys.\ Rev.\ D48 (1993) 3160.
\item\label{bcmn} 
  R. Bonciani, S. Catani, M.L. Mangano and P. Nason,
  Nucl.\ Phys.\ B529 (1998) 424.
\item\label{cdf} 
  The CDF II Detector, Technical Design Report,
  FERMILAB-Pub-96/390-E.
\item\label{atlas} 
  ATLAS detector and physics performance, Technical
  Design Report, Volume I, CERN LHCC 99-14.
\item\label{SLD}
 K.~Abe {\it et al.}  [SLD Collaboration],
  hep-ex/9912058.
\end{enumerate}
\end{document}